\documentclass[trackchanges,twocolumn]{aastex7}

\newcommand{\kkoname}
{k'ni\textipa{P}atn k'l$\left._\mathrm{\smile}\right.$stk'masqt}

\newcommand{\kkonamecaps}{K'ni\textipa{P}atn k'l$\left._\mathrm{\smile}\right.$stk'masqt}
\usepackage{tipa}
\usepackage{gensymb}
\usepackage{amsmath,amstext}
\usepackage{amssymb}
\usepackage{CJKutf8} 
\begin{document}
\begin{CJK*}{UTF8}{gbsn}

\title{Discovery and Localization of the Swift-Observed FRB 20241228A in a Star-forming Host Galaxy}

\author[0000-0002-8376-1563]{Alice Curtin}
  \email{alice.curtin@mail.mcgill.ca}
  \affiliation{Department of Physics, McGill University, 3600 rue University, Montr\'eal, QC H3A 2T8, Canada}
  \affiliation{Trottier Space Institute, McGill University, 3550 rue University, Montr\'eal, QC H3A 2A7, Canada}
\author[0000-0002-3980-815X]{Shion Andrew}
  \email{shiona@mit.edu}
  \affiliation{MIT Kavli Institute for Astrophysics and Space Research, Massachusetts Institute of Technology, 77 Massachusetts Ave, Cambridge, MA 02139, USA}
  \affiliation{Department of Physics, Massachusetts Institute of Technology, 77 Massachusetts Ave, Cambridge, MA 02139, USA}
\author[0000-0003-3801-1496]{Sunil Simha}
  \email{sunil.simha@northwestern.edu}
  \affiliation{Center for Interdisciplinary Exploration and Research in Astronomy, Northwestern University, 1800 Sherman Avenue, Evanston, IL 60201, USA }
  \affiliation{Department of Astronomy and Astrophysics, University of Chicago, William Eckhardt Research Center, 5640 S Ellis Ave, Chicago, IL 60637, USA}
\author[0009-0001-0983-623X]{Alice Cai}
  \email{acai@u.northwestern.edu}
  \affiliation{Center for Interdisciplinary Exploration and Research in Astronomy, Northwestern University, 1800 Sherman Avenue, Evanston, IL 60201, USA }
\author[0000-0003-0510-0740]{Kenzie Nimmo}
  \email{knimmo@mit.edu}
  \affiliation{MIT Kavli Institute for Astrophysics and Space Research, Massachusetts Institute of Technology, 77 Massachusetts Ave, Cambridge, MA 02139, USA}
\author[0000-0002-2878-1502]{Shami Chatterjee}
  \email{shami@astro.cornell.edu}
  \affiliation{Cornell Center for Astrophysics and Planetary Science, Cornell University, Ithaca, NY 14853, USA}
\author[0000-0001-6422-8125]{Amanda M.~Cook}
  \email{amanda.cook@mcgill.ca}
  \affiliation{Department of Physics, McGill University, 3600 rue University, Montr\'eal, QC H3A 2T8, Canada}
  \affiliation{Trottier Space Institute, McGill University, 3550 rue University, Montr\'eal, QC H3A 2A7, Canada}
  \affiliation{Anton Pannekoek Institute for Astronomy, University of Amsterdam, Science Park 904, 1098 XH Amsterdam, The Netherlands}
\author[0000-0003-4098-5222]{Fengqiu Adam Dong}
  \email{fengqiu.dong@gmail.com}
  \affiliation{National Radio Astronomy Observatory, 520 Edgemont Rd, Charlottesville, VA 22903, USA}
\author[0000-0002-9363-8606]{Yuxin Dong}
  \email{yuxin.dong@northwestern.edu}
  \affiliation{Center for Interdisciplinary Exploration and Research in Astronomy, Northwestern University, 1800 Sherman Avenue, Evanston, IL 60201, USA }
\author[0000-0003-0307-9984]{Tarraneh Eftekhari}
  \email{teftekhari@northwestern.edu}
  \affiliation{Center for Interdisciplinary Exploration and Research in Astronomy, Northwestern University, 1800 Sherman Avenue, Evanston, IL 60201, USA }
\author[0000-0002-7374-935X]{Wen-fai Fong}
  \email{wfong@northwestern.edu}
  \affiliation{Center for Interdisciplinary Exploration and Research in Astronomy, Northwestern University, 1800 Sherman Avenue, Evanston, IL 60201, USA }
\author[0000-0001-8384-5049]{Emmanuel Fonseca}
  \email{emmanuel.fonseca@mail.wvu.edu}
  \affiliation{Department of Physics and Astronomy, West Virginia University, PO Box 6315, Morgantown, WV 26506, USA }
  \affiliation{Center for Gravitational Waves and Cosmology, West Virginia University, Chestnut Ridge Research Building, Morgantown, WV 26505, USA}
\author[0000-0003-2317-1446]{Jason W.~T.~Hessels}
  \email{jason.hessels@mcgill.ca}
  \affiliation{Department of Physics, McGill University, 3600 rue University, Montr\'eal, QC H3A 2T8, Canada}
  \affiliation{Trottier Space Institute, McGill University, 3550 rue University, Montr\'eal, QC H3A 2A7, Canada}
  \affiliation{Anton Pannekoek Institute for Astronomy, University of Amsterdam, Science Park 904, 1098 XH Amsterdam, The Netherlands}
  \affiliation{ASTRON, Netherlands Institute for Radio Astronomy, Oude Hoogeveensedijk 4, 7991 PD Dwingeloo, The Netherlands}
\author[0000-0003-3457-4670]{Ronniy C.~Joseph}
  \email{ronniy.joseph@mcgill.ca}
  \affiliation{Department of Physics, McGill University, 3600 rue University, Montr\'eal, QC H3A 2T8, Canada}
  \affiliation{Trottier Space Institute, McGill University, 3550 rue University, Montr\'eal, QC H3A 2A7, Canada}
\author[0000-0001-9345-0307]{Victoria Kaspi}
  \email{victoria.kaspi@mcgill.ca}
  \affiliation{Department of Physics, McGill University, 3600 rue University, Montr\'eal, QC H3A 2T8, Canada}
  \affiliation{Trottier Space Institute, McGill University, 3550 rue University, Montr\'eal, QC H3A 2A7, Canada}
\author[0000-0002-4209-7408]{Calvin Leung}
  \email{calvin_leung@berkeley.edu}
  \affiliation{Department of Astronomy, University of California, Berkeley, CA 94720, United States}
  \affiliation{Miller Institute for Basic Research, Stanley Hall, Room 206B, Berkeley, CA 94720, USA}
\author[0000-0002-7164-9507]{Robert Main}
  \email{robert.main@mcgill.ca}
  \affiliation{Department of Physics, McGill University, 3600 rue University, Montr\'eal, QC H3A 2T8, Canada}
  \affiliation{Trottier Space Institute, McGill University, 3550 rue University, Montr\'eal, QC H3A 2A7, Canada}
\author[0000-0002-4279-6946]{Kiyoshi W.~Masui}
  \email{kmasui@mit.edu}
  \affiliation{MIT Kavli Institute for Astrophysics and Space Research, Massachusetts Institute of Technology, 77 Massachusetts Ave, Cambridge, MA 02139, USA}
  \affiliation{Department of Physics, Massachusetts Institute of Technology, 77 Massachusetts Ave, Cambridge, MA 02139, USA}
\author[0000-0001-7348-6900]{Ryan Mckinven}
  \email{ryan.mckinven@mcgill.ca}
  \affiliation{Department of Physics, McGill University, 3600 rue University, Montr\'eal, QC H3A 2T8, Canada}
  \affiliation{Trottier Space Institute, McGill University, 3550 rue University, Montr\'eal, QC H3A 2A7, Canada}
\author[0000-0002-2551-7554]{Daniele Michilli}
  \email{danielemichilli@gmail.com}
  \affiliation{Laboratoire d'Astrophysique de Marseille, Aix-Marseille Univ., CNRS, CNES, Marseille, France}
\author[0000-0002-0940-6563]{Mason Ng}
  \email{mason.ng@mcgill.ca}
  \affiliation{Department of Physics, McGill University, 3600 rue University, Montr\'eal, QC H3A 2T8, Canada}
  \affiliation{Trottier Space Institute, McGill University, 3550 rue University, Montr\'eal, QC H3A 2A7, Canada}
\author[0000-0002-8897-1973]{Ayush Pandhi}
  \email{ayush.pandhi@mail.utoronto.ca}
  \affiliation{David A.\ Dunlap Department of Astronomy and Astrophysics, 50 St. George Street, University of Toronto, ON M5S 3H4, Canada}
  \affiliation{Dunlap Institute for Astronomy and Astrophysics, 50 St. George Street, University of Toronto, ON M5S 3H4, Canada}
\author[0000-0002-8912-0732]{Aaron B.~Pearlman}
  \email{aaron.b.pearlman@physics.mcgill.ca}
  \affiliation{Department of Physics, McGill University, 3600 rue University, Montr\'eal, QC H3A 2T8, Canada}
  \affiliation{Trottier Space Institute, McGill University, 3550 rue University, Montr\'eal, QC H3A 2A7, Canada}

\author[0000-0002-4795-697X]{Ziggy Pleunis}
    \email{z.pleunis@uva.nl}
   \affiliation{Anton Pannekoek Institute for Astronomy, University of Amsterdam, Science Park 904, 1098 XH Amsterdam, The Netherlands}
  \affiliation{ASTRON, Netherlands Institute for Radio Astronomy, Oude Hoogeveensedijk 4, 7991 PD Dwingeloo, The Netherlands}
\author[0000-0002-4623-5329]{Mawson W.~Sammons}
  \email{mawson.sammons@mcgill.ca}
  \affiliation{Department of Physics, McGill University, 3600 rue University, Montr\'eal, QC H3A 2T8, Canada}
  \affiliation{Trottier Space Institute, McGill University, 3550 rue University, Montr\'eal, QC H3A 2A7, Canada}
\author[0000-0003-3154-3676]{Ketan R Sand}
  \email{ketan.sand@mail.mcgill.ca}
  \affiliation{Department of Physics, McGill University, 3600 rue University, Montr\'eal, QC H3A 2T8, Canada}
  \affiliation{Trottier Space Institute, McGill University, 3550 rue University, Montr\'eal, QC H3A 2A7, Canada}
\author[0000-0002-7374-7119]{Paul Scholz}
  \email{pscholz@yorku.ca}
  \affiliation{Department of Physics and Astronomy, York University, 4700 Keele Street, Toronto, ON MJ3 1P3, Canada}
  \affiliation{Dunlap Institute for Astronomy and Astrophysics, 50 St. George Street, University of Toronto, ON M5S 3H4, Canada}
\author[0000-0002-4823-1946]{Vishwangi Shah}
  \email{vishwangi.shah@mail.mcgill.ca}
  \affiliation{Department of Physics, McGill University, 3600 rue University, Montr\'eal, QC H3A 2T8, Canada}
  \affiliation{Trottier Space Institute, McGill University, 3550 rue University, Montr\'eal, QC H3A 2A7, Canada}
\author[0000-0002-6823-2073]{Kaitlyn Shin}
  \email{kshin@mit.edu}
  \affiliation{MIT Kavli Institute for Astrophysics and Space Research, Massachusetts Institute of Technology, 77 Massachusetts Ave, Cambridge, MA 02139, USA}
  \affiliation{Department of Physics, Massachusetts Institute of Technology, 77 Massachusetts Ave, Cambridge, MA 02139, USA}
\author[0000-0002-2810-8764]{Aaron Tohuvavohu}
  \email{aaron.tohu@gmail.com}
  \affiliation{Division of Physics, Mathematics, and Astronomy, California Institute of Technology, Pasadena, CA 91125, USA}
\newcommand{\allacks}{
Acknowledgements
A.P.C. is a Vanier Canada Graduate Scholar. 
S.S. is supported by the joint Northwestern University and University of Chicago Brinson Fellowship.
K.N. is an MIT Kavli Fellow.
A.M.C. is a Banting Postdoctoral Researcher.
F.A.D. is supported by the Jansky Fellowship.
Y.D. is supported by the National Science Foundation Graduate Research Fellowship under grant No. DGE-2234667. 
W.F. gratefully acknowledges support by the David and Lucile Packard Foundation, the Research Corporation for Science Advancement through Cottrell Scholar Award \#28284, and the NSF (AST-2206494, CAREER grant AST-2047919).
E.F. is supported by the National Science Foundation (NSF) under grant number AST-2407399.
J.W.T.H. and the AstroFlash research group acknowledge support from a Canada Excellence Research Chair in Transient Astrophysics (CERC-2022-00009); an Advanced Grant from the European Research Council (ERC) under the European Union’s Horizon 2020 research and innovation programme (`EuroFlash'; Grant agreement No. 101098079); and an NWO-Vici grant (`AstroFlash'; VI.C.192.045).
V.M.K. holds the Lorne Trottier Chair in Astrophysics \& Cosmology, a Distinguished James McGill Professorship, and receives support from an NSERC Discovery grant (RGPIN 228738-13).
C. L. acknowledges support from the Miller Institute for Basic Research at UC Berkeley.
K.W.M. holds the Adam J. Burgasser Chair in Astrophysics and is supported by NSF grant 2018490.
D.M. acknowledges support from the French government under the France 2030 investment plan, as part of the Initiative d'Excellence d'Aix-Marseille Universit\'e -- A*MIDEX (AMX-23-CEI-088).
M.N. is a Fonds de Recherche du Quebec -- Nature et Technologies~(FRQNT) postdoctoral fellow.
A.P. is funded by the NSERC Canada Graduate Scholarships -- Doctoral program.
A.B.P. is a Banting Fellow, a McGill Space Institute~(MSI) Fellow, and a Fonds de Recherche du Quebec -- Nature et Technologies~(FRQNT) postdoctoral fellow.
Z.P. is supported by an NWO Veni fellowship (VI.Veni.222.295).
M.W.S. acknowledges support from the Trottier Space Institute Fellowship program.
K.R.S is supported by FRQNT doctoral research award
P.S. acknowledges the support of an NSERC Discovery Grant (RGPIN-2024-06266).
V.S. is supported by FRQNT doctoral research award
K.S. is supported by the NSF Graduate Research Fellowship Program.
}

\correspondingauthor{Alice P. Curtin}
\email{alice.curtin@mail.mcgill.ca}

\shortauthors{Curtin et al.}
\begin{abstract}

On 2024 December 28, CHIME/FRB detected the thus-far non-repeating FRB 20241228A with a real-time signal-to-noise ratio of $>50$. Approximately 112~s later, the X-ray Telescope onboard the {\it Neil Gehrels Swift Observatory} was on source, the fastest follow-up to-date of a non-repeating FRB (Tohuvavohu et al. \textit{in prep.}). Using CHIME/FRB and two of the three CHIME/FRB Outriggers, we obtained a Very Long Baseline Interferometry localization for FRB 20241228A with a 1$\sigma$ confidence ellipse of 11$^{\prime\prime}$ by 0.2$^{\prime\prime}$. This represents the first published localization using both the CHIME-\kkoname{} and CHIME-Green Bank Outriggers. We associate FRB 20241228A with a star-forming galaxy at a redshift of $z = 0.1614\pm0.0002$. The persistent X-ray luminosity limit at this source's location and distance is $<1.2 \times 10^{43}$ erg s$^{-1}$ in the $0.3-10$ keV band, the most stringent limit of any non-repeating FRB to-date (Tohuvavohu et al. \textit{in prep.}). The stellar mass ($\sim 2.6 \times 10^{10}\,M_{\odot}$) and star formation rate ($\sim 2.9\,M_{\odot}$~yr$^{-1}$) of the host galaxy of FRB 20241228A are consistent with the broader FRB host galaxy population. We measure significant scattering ($\sim$2ms) and scintillation ($\sim$20 kHz at 600 MHz) along the line of sight to this source, and suggest the scintillation screen is Galactic while the scattering screen is extragalactic. FRB 20241228A represents an exciting example of a new era in which we can harness VLBI-localizations and rapid high-energy follow-up to probe FRB progenitors.

\end{abstract}

\keywords{\uat{Radio transient sources}{2008} --- \uat{Radio Astronomy}{1338} --- \uat{Very long baseline interferometry}{1769} }

\section{Introduction}
Searching for high-energy counterparts to fast radio bursts (FRBs) is a promising avenue for constraining the nature of FRB progenitors.  In April 2020, the detection of an FRB-like burst associated with the Galactic magnetar SGR 1935+2154 strongly suggested that at least some FRBs arise from magnetars \citep{abb+20, brb+20}. Interestingly, the radio burst from SGR 1935+2154 occurred at a time of increased X-ray activity \citep{pal20}, and was accompanied by simultaneous high-energy emission in both the X-rays and soft gamma-rays \citep{2021NatureInsightSGR, rsf+21, msf+20}. 

Observing similar high-energy emission from an extragalactic FRB would further support the connection between FRBs and magnetars. The detection would also provide critical insight into the total energy budget of FRBs, and might point to a connection with other classes of transients from similar progenitors e.g., supernova explosions, gamma-ray bursts, and accreting X-ray binaries. However, no FRBs have been detected from other known Galactic magnetars \citep{tkp16} or proposed magnetar flares \citep{2023ATel16341....1C}. Furthermore, despite many follow-up studies of FRB sites and cross-matching with known high-energy transients, there has been no significant detection or association of high-energy emission with an extragalactic FRB (e.g., in optical: \citealt{kilpatrick,2017MNRAS.472.2800H, 2023ApJ...947L..28H}, in X-ray: \citealt{sbh+17,scc+20,Pilia_2020, 2024arXiv240212084Y,2023ApJ...958...66E,Cook_2024,2025NatAs...9..111P}, in gamma-ray: \citealt{2016MNRAS.460.2875Y,2020ApJ...890L..32C, ctj+23, csk+24}). For FRB 20201124A, \cite{2021AnA...656L..15P} reported the detection of a faint, persistent X-ray source spatially coincident with the location of the FRB, but they concluded that the emission was consistent with that expected from star-formation in the region. 

Two key challenges in detecting high-energy emission from extragalactic FRBs are the small number of known repeaters and the stochastic nature of their activity. Only $\sim3\%$ of FRBs have been seen to repeat \citep{abb+23}, and of those, only one (FRB 20180916B) repeats with a definitive activity period \citep{aab+20}. Thus, the coordination of high-energy observations at the time of a radio detection is extremely difficult. Some repeating sources show periods of high source activity, which have proved fruitful for high-energy follow-up with X-ray telescopes \citep{sbh+17, Pilia_2020, scc+20, Trudu+2023, Cook_2024, 2025NatAs...9..111P}. 
However, for non-repeating FRBs, similar observations have relied on serendipitous observations with wide-field X-ray telescopes at the times of FRBs \citep{Anumarlapudi+2020, Guidorzi+2020, Waratkar+2025}, making simultaneous observations infeasible in most cases. Instead, extremely rapid follow-up observations are often required to be able to capture near-simultaneous multi-wavelength emission.

The Canadian Hydrogen Intensity Mapping Experiment (CHIME)/FRB project has detected nearly 5000 FRBs since its initial commissioning in 2018 (CHIME/FRB Collaboration \textit{submitted}), with a daily FRB detection rate of $2-3$ FRBs per day \citep{abb+18}. CHIME/FRB is equipped with a VOEvent service to provide near-immediate ($\sim$10 s) alerts after an FRB \citep{phb+17, 2025AJ....169...39A},  key for rapid multi-wavelength follow-up. This capability enables real-time coordination with telescopes like the {\it Neil Gehrels Swift Observatory} \citep{Gehrels2004}, which has an automatic re-pointing algorithm to ingest triggers such as CHIME/FRB's VOEvents \citep{2020ApJ...900...35T}.

On 2024 December 28, CHIME/FRB detected a bright FRB with a real-time signal-to-noise (S/N) ratio of 50. The X-ray telescope (XRT) onboard \textit{Swift} immediately slewed to the source and was on target collecting photons within 112 seconds. Offline analysis determined an improved localization for the FRB\footnote{Due to CHIME's strong sidelobes, real-time FRB localizations can span $\sim5 - 10$~degrees in right ascension (RA) and $\sim1$ degree in declination (DEC).}, and confirmed that the source was located within \textit{Swift}/XRT's field of view (FOV).  Tohavavohu et al. \textit{in prep.} present a detailed description of the \textit{Swift} triggering process, as well as the persistent X-ray limit at the position of FRB 20241228A. Here, we present the CHIME/Outriggers \citep{chimeOutriggers} Very Long Baseline Interferometry (VLBI) localization for FRB 20241228A along with the most probable host galaxy and redshift. This is the first FRB localization to include two of the three CHIME/FRB Outriggers, including the longest Outrigger baseline. We also discuss the burst morphology, with particular focus on the scintillation and scattering features present in the burst, and the non-repeating nature of the source. 

Below, in Section \ref{sec:observations}, we introduce CHIME/FRB, the CHIME/FRB Outriggers, and present the VLBI localization. In Section \ref{sec: Host galaxy}, we discuss the most probable host galaxy associated with FRB 20241228A. We then discuss the burst morphology, and in particular, a detection of scintillation within FRB 20241228A in Section \ref{sec: burst properties}. We conclude in Section \ref{sec: Discussion} by discussing the host galaxy, repetition rate, and scintillation features of the burst in the context of the broader FRB field. 

\section{Observations} 
\label{sec:observations}
\subsection{CHIME/FRB Burst Discovery}

\noindent CHIME is located near Penticton, British Columbia at the Dominion Radio Astrophysical Observatory (DRAO). Consisting of four, 100-m by 20-m cylindrical, parabolic dishes, CHIME has a total FOV of $\sim$250 deg$^2$ \citep[dependent on frequency; ][]{nvp+17}. Each dish consists of 256 dual polarization antennas operating between 400 and 800 MHz. FRB-like signals are searched for in total intensity data at a time resolution of 0.983 ms. For events with a S/N ratio greater than a specified cutoff (typically $\approx$ 12), we save the channelized raw voltage data (1024 frequency channelsl) at a time resolution of 2.56 $\mu$s. This allows for both improved localization of the event through offline beamforming \citep[of order arc-minutes; often referred to as a `baseband' localization; ][]{mmm+21, aaa+24} as well as studies of the burst's polarization and morphology properties. 

\begin{figure}
    \centering
    {\includegraphics[width=\columnwidth]{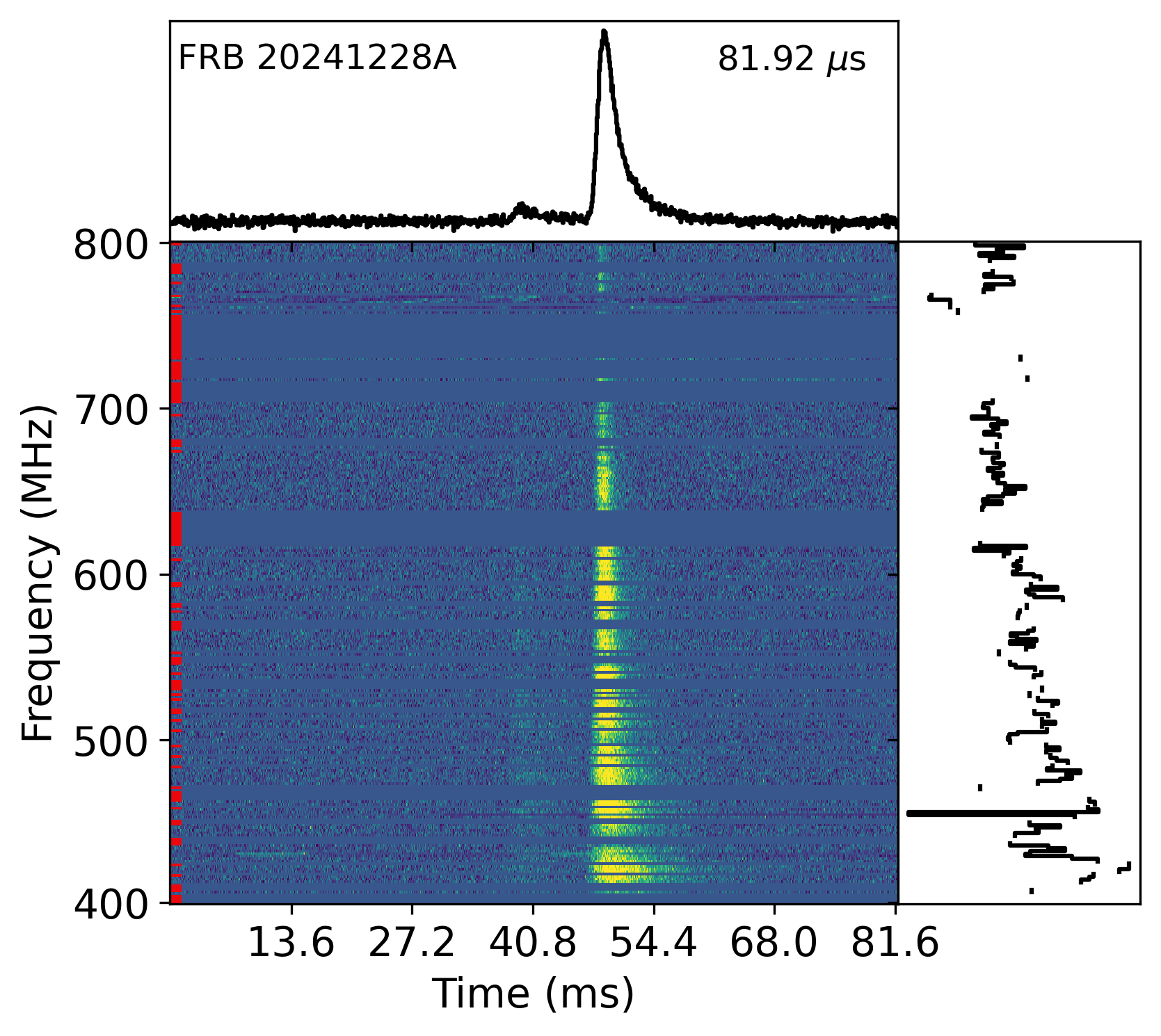} }
    \caption{Dynamic spectrum for FRB 20241228A at a time resolution of  81.92 $\mu$s using the raw voltage data. The time is relative to the fiducial arrival time of the burst. The data are scrunched in frequency by a factor of 4, for a resolution of 1.56 MHz. The top panel shows the frequency-summed time series while the right panel shows the burst's spectrum averaged over both the burst and the surrounding noise.  The $\sim$30-MHz ripple seen in the spectrum is not astrophysical and instead corresponds to reflections between the cylinders and the focal lines. Channels which are either flagged due to RFI or which are missing due to networking issues are indicated with red ticks. }
    \label{fig:waterfalls}
\end{figure}

FRB 20241228A was detected by CHIME/FRB's realtime pipeline on 2024 December 28 at 15:55:56 UTC (topocentric at CHIME near Penticton, Canada) with a real-time S/N of 50.2 (see Figure \ref{fig:waterfalls}). The real-time pipeline dispersion measure (DM) of the source was $246.3 \pm 0.4$ pc cm$^{-3}$. The maximum Galactic DM contribution in the direction of the event is  23.8 pc cm$^{-3}$ in the NE2001 electron density model \citep{ne2001} or  22.75 pc cm$^{-3}$ in the YMW17 model \citep{ymw17}, significantly below the source's DM and hence confirming its extragalactic nature. Despite 119 hours of exposure at this source's location (see Section \ref{subsec: rate}), we have not yet observed any other candidate bursts from this source, i.e., any other events in our database which are (i) consistent on the sky with the position of FRB 20241228A, (ii) within 5 pc cm$^{-3}$ DM of the DM and (iii) above our nominal S/N threshold of 8. CHIME/FRB has some sensitivity to very bright bursts in our sidelobes, i.e., we can detect bright bursts like Crab pulsar giant pulses many degrees off meridian \citep[e.g.][]{lsn+23}. We perform the same database search along the apparent positional arc of FRB 20241228A and find no plausible sidelobe repeats of the source. In short, FRB 20241228A is a thus-far non-repeater.

\subsection{CHIME/FRB Outriggers Localizations}

\noindent The CHIME/FRB Outriggers (herein referred to as the Outriggers) are a VLBI network of three, CHIME-like cylindrical telescopes located across North America. With a maximum baseline of 3370 km, the Outriggers will localize hundreds to thousands of CHIME-detected FRBs to $\sim$50-mas precision \citep{chimeOutriggers}. The first Outrigger, \kkoname{} (KKO)\footnote{The name of the first Outrigger \kkoname{} was a generous gift from the
Upper Similkameen Indian Band and means ``a listening device for outer space.''}, is located in British Columbia, 66-km west of CHIME, and provides improved localizations of order an arcsecond in RA \citep{lanman_kko}. The KKO Outrigger has already localized and identified the host galaxies for 21 FRBs \citep{ amiri2025cataloglocaluniversefast}, and was the first to localize a repeating FRB to the outskirts of a quiescent galaxy \citep{ ssl+24, edf+24}.  

The second CHIME/Outrigger is located at the National Radio Astronomy Observatory in Green Bank, West Virginia and, in conjunction with the CHIME-KKO baseline, provides 50 mas by 20 arcsec localization capabilities. Because the CHIME-GBO baseline is nearly co-linear with the CHIME-KKO baseline \citep[see also Figure 2 of ][]{chimeOutriggers}, a third Outrigger located in Hat Creek, California at the Hat Creek Radio Observatory (HCRO) is required to provide the u-v coverage necessary for a full 2D localization of $\sim 50 \times$ 150 mas. The Hat Creek Outrigger is presently under commissioning. 

At the time of FRB 20241228A, both the KKO and GBO Outrigger stations were operating in full scientific mode. The HCO Outrigger was still undergoing final digital and analog deployment, and hence was not online. With a CHIME/FRB detection S/N greater than 15, FRB 20241228A triggered raw voltage data capture at CHIME, the KKO Outrigger, and the GBO Outrigger. FRB 20241228A thus represents the first localization using both the CHIME-KKO and CHIME-GBO baselines. Description of the CHIME-KKO system, and its localization capabilities, have already been discussed in detail in \citet{lanman_kko}. Below, we provide a brief overview of the CHIME-GBO system and its astrometric capabilities. An in-depth description of the construction, calibration, and validation of the GBO Outrigger will be discussed in detail in a future paper.

The calibration and validation strategy for the GBO Outrigger has closely matched that for the KKO Outrigger. Bright, in-beam sources identified in our calibrator survey~\citep{andrew2024vlbicalibratorgrid600mhz} are observed commensally  as delay calibrators for each of our long baselines \citep{leung_2021}. For KKO-detected FRBs, there are approximately 10 in-beam calibrators in a given FRB voltage dump. However, given the significantly longer baseline of GBO as compared to KKO, this number reduces to only two in-beam calibrators as many are resolved out in the longer baseline and/or are too faint to be detected.  While GBO has a significantly larger collecting area KKO, the S/N is reduced along its baseline due to significant RFI at site. A more detailed description of the S/N reduction on the CHIME-GBO baseline will be provided in future work.  

We derive our error budget for this source from test localizations of $\sim$80 pulsars localized to mas-level precision \citep[VLBA/21A-314, PI: Kaczmarek; VLBA/22A-345, PI: Curtin; VLBA/23A-099, PI: Curtin; VLBA/24B-328, PI: Curtin;][]{Walter_Brisken,Lin_2023,Chatterjee_2009,Deller_2019} and over 200 International Celestial Reference Frame calibrators. For test localizations with similar target-calibrator separations (discussed below), the rms localization error on the CHIME-GBO baseline is smaller than 200~mas. Thus, we provide a final uncertainty of 200-mas along the CHIME-GBO baseline for this FRB’s localization region. This error budget will be discussed further in future work.

We used the \texttt{pyfx} software \citep{Leung_2024} to form correlated visibilities between CHIME and the KKO and GBO Outriggers. FRB 20241228A was initially detected by CHIME/FRB with a real-time (at a time resolution of 0.983 ms) S/N of $\sim$50. After re-beamforming the data, the S/N is $>200$ at a time resolution of 0.655 ms. It was thus detected in the cross correlated visibilities with a S/N of $\sim$ 45 along the CHIME-KKO baseline and with a S/N of $\sim$ 35 along the CHIME-GBO baseline. Two in-beam calibrators were detected in the voltage dumps at both KKO and GBO. For both baselines, we chose J143844.7+621154, the closer of the two, as our calibrator. Located 50\degree \space from FRB 20241228A, J143844.7+621154 was detected with a cross-correlation S/N of 30.

\begin{figure}
    \centering
    {\includegraphics[width=\columnwidth]{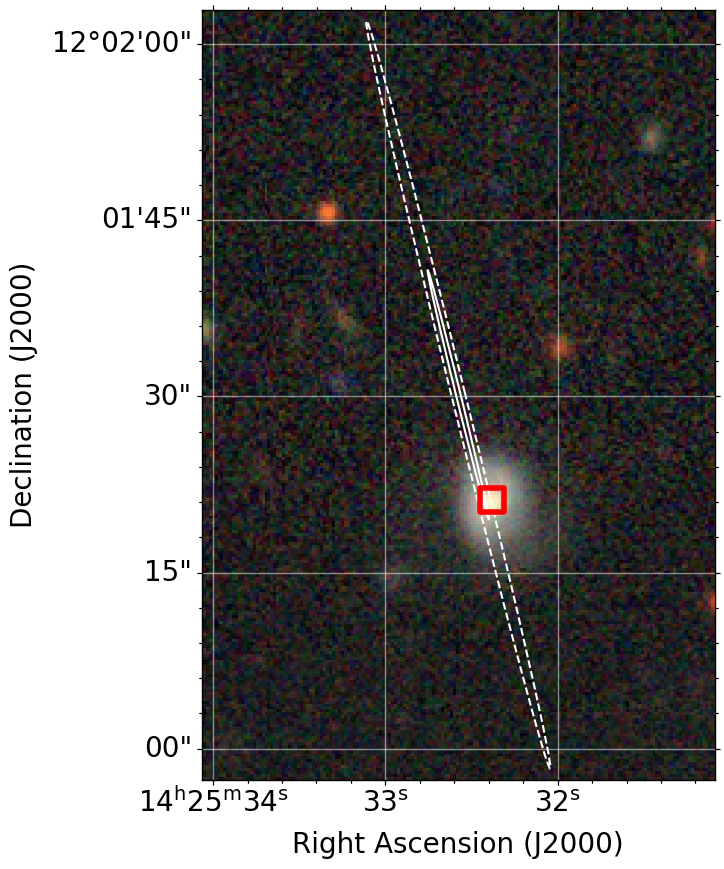} }
    \caption{Joint localization (CHIME-GBO, CHIME-KKO, and CHIME/FRB baseband) on top of the DECaLS field for FRB 20241228A. The 1-$\sigma$  localization is shown with a white, solid line while the 3-$\sigma$ ellipse is shown with a white, dashed line. The most probable host galaxy (P(O$|$x) = 0.96) is shown with a red square.}
    \label{fig:localization}
\end{figure}

\begin{figure}
    \centering
    \includegraphics[width=0.85\columnwidth]{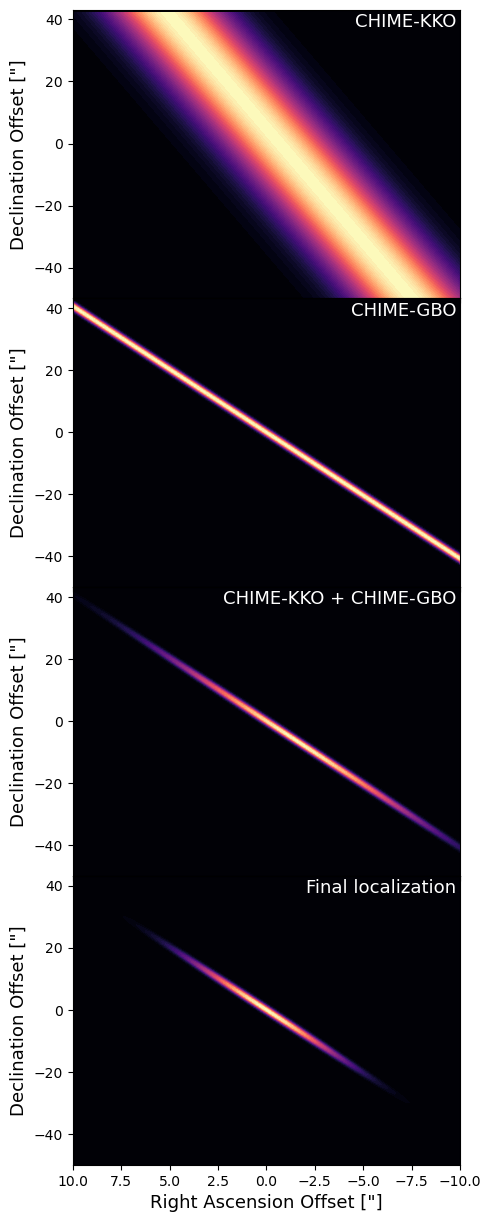}
    \caption{\textit{Top panel:} Probability map for the localization of FRB 20241228A using solely the CHIME-KKO baseline. The offsets in RA and DEC are relative to the center of the final localization ellipse. Lighter colors represent areas with higher localization probability. 
    \textit{Middle top panel:} Same as first except for the CHIME-GBO baseline. 
    \textit{Middle lower panel: } Combined localization using the CHIME-KKO and CHIME-GBO baselines. Due to the near co-linear baselines of CHIME-KKO and CHIME-GBO, the resolving power from combining the two is largely along a single axis. 
    \textit{Bottom panel: } Combined localization using the CHIME-KKO baseline, CHIME-GBO baseline, and CHIME/FRB baseband localization. The baseband data significantly constrains the final localization in declination. }
    \label{fig:KKOGBOCombinedLocalizations}
\end{figure}

In Figure \ref{fig:localization}, we show the $1\sigma$ and 3$\sigma$ localization ellipses from CHIME-GBO baseline overlaid on a Dark Energy Camera Legacy Survey \citep[DECaLS;][]{dsl+19} image at this location.  The parameters for the ellipse are given in Table \ref{tab:burst_loc}. As shown in Figure \ref{fig:KKOGBOCombinedLocalizations}, the CHIME-KKO and CHIME-GBO baselines each provide 1D localizations that are primarily constrained along the E-W direction. Thus, the final uncertainty in the semi-minor axis of the combined CHIME-KKO-GBO localization is significantly smaller than that of the semi-major axis. We note that for the final ellipse region (bottom panel of Figure \ref{fig:KKOGBOCombinedLocalizations}), we have also incorporated CHIME/FRB's baseband localization, which significantly constrains the localization in declination.

\begin{deluxetable*}{ c c c c c }
\tabletypesize{\normalsize}
\tablecaption{Burst Localization
\label{tab:burst_loc}}
\tablehead{
\colhead{Right Ascension} & \colhead{Declination} & \colhead{Semi-minor Axis} & \colhead{Semi-major Axis} & \colhead{Position Angle} }
 \startdata
{${14{\mathrm{h}}25{\mathrm{m}}32.58{\mathrm{s}}}$} & {${12^{\circ}01^{\prime}30.15^{\prime\prime}}$} & {10.93$^{\prime\prime}$} & {0.20$^{\prime\prime}$} & {$13.805527^{\circ}$}
\enddata
\tablecomments{Parameters for the 1-$\sigma$ CHIME/Outriggers VLBI localization ellipse for FRB 20241228A, with the center of the ellipse defined in the ICRS J2000 reference frame. \label{tab:VLBI_pos}}
\end{deluxetable*}

\section{Host Galaxy Localization and Association} 
\label{sec: Host galaxy}
\subsection{Host association}
To determine the most probable host galaxy of FRB 20241228A, we used the   Probabilistic Association of Transients to their Hosts (PATH) framework  (\citealt{aba+21}). We utilize DECaLS to identify host galaxy candidates in the field of FRB 20241228A down to a $5\sigma$ limiting $r$-band magnitude $m_r = 23.4$ mag \citep{Dey+2019}. We adopt the default ``inverse" prior to account for the larger number of faint galaxies and a prior probability that the host is unseen of P(O$|$x) = 0.15, motivated by the simulations of Andersen et al. in prep.  For the prior on the FRB offset distribution with respect to its host, we assume an underlying exponential distribution with a scale length equal to one half the galaxy half-light radius. Using the FRB coordinates and ellipse parameters presented in Table \ref{tab:VLBI_pos}, we identified the most likely host galaxy to have a posterior probability $P(O|x) = 0.96$. It is the only galaxy apparent in DECaLS that intersects the Outrigger position (see Figure \ref{fig:localization}). We obtain a position for the host galaxy of J142532.4s+120121.2.

\subsection{Spectroscopic Observations}
Using the Gemini Multi-Object Spectrograph (GMOS) mounted on the 8-meter Gemini-North telescope, we obtained optical spectroscopy of the host of FRB 20241228A on 2025 March 3 UT (PI: T.~Eftekhari)\footnote{Spectroscopic data were collected by the Fast and Fortunate for FRB Follow-up Collaboration (\url{https://www.frb-f4.org/}) using the \texttt{FFFF-PZ} observation management tool, built and modeled after \texttt{YSE-PZ} \citep{Coulter2022,Coulter2023}, and first implemented for FRBs in \citet{amiri2025cataloglocaluniversefast}.}. We took a total of $4 \times 900$ s exposures with a 1\arcsec\ slit width using the B480 grating and the GG455 blocking filter at central wavelengths of 640 and 650 nm. We reduced and coadded the spectra using the Python Spectroscopic Data Reduction Pipeline \citep[\texttt{PypeIt}, v1.16;][]{pypit}. We applied absolute flux calibration and a telluric correction to the spectra using a spectrophotometric standard star observation taken in the same spectral set-ups, and atmospheric model grids available in \texttt{PypeIt}.

\begin{figure*}
    \centering
    {\includegraphics[width=0.95\textwidth]{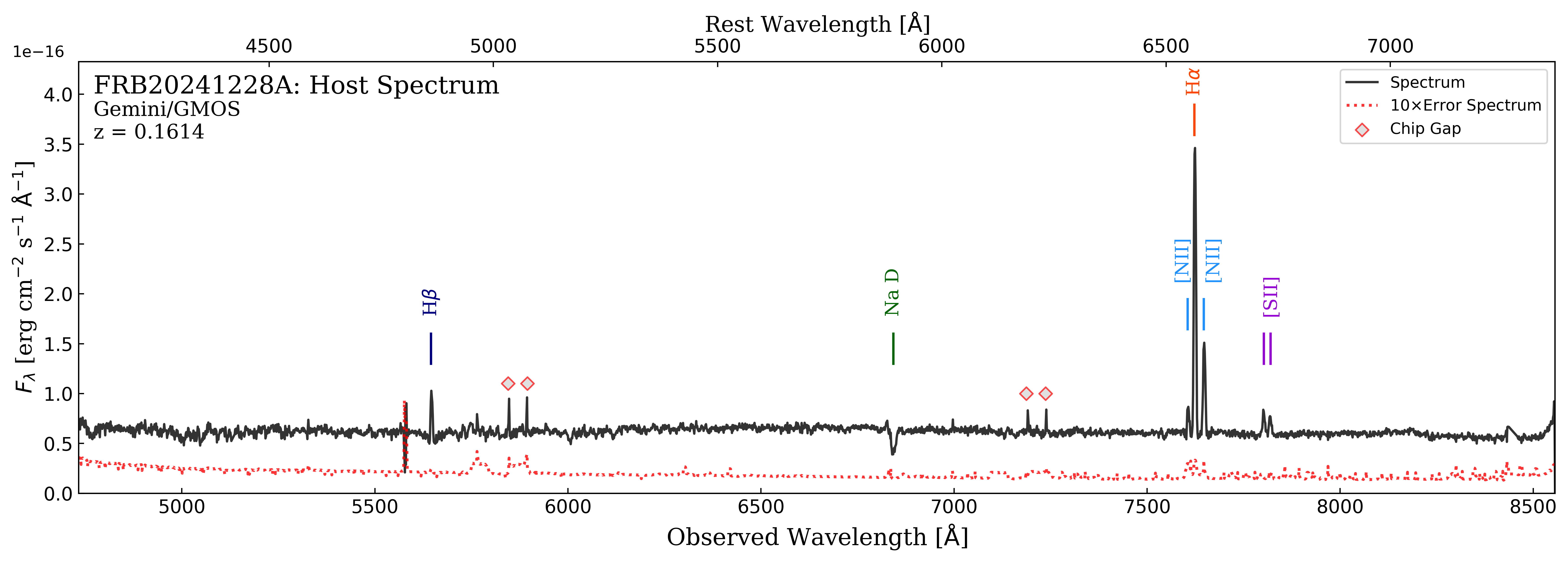} }
    \caption{Gemini/GMOS spectrum of the most probable host galaxy of FRB 20241228A (black). Also plotted is the error spectrum (red dotted), which has been scaled by a factor of 10 for readability. The locations of observed spectral features which facilitated a redshift determination are indicated. False features induced by the detector chip gaps are marked by the red-bordered diamonds.
    }
    \label{fig:HostGalSpectrum}
\end{figure*}

To determine the redshift of the galaxy, we utilized the Manual and Automatic Redshifting (\texttt{Marz}) analysis software \citep{Hinton_2016}. \texttt{Marz} performs a cross-correlation between the measured spectra and various archetypal template models\footnote{We used the six standard galaxy templates for this cross-correlation.} to determine both the broad galaxy type (e.g., dominated by emission lines or absorption features) and its redshift. Applying \texttt{Marz} to the host galaxy of FRB 20241228A yields a redshift of $z = 0.1614\pm0.0002$. \texttt{Marz} does not report a redshift uncertainty but we derive this from the $1\sigma$ width of the lines.  The prominent spectral emission features (H$\alpha$, H$\beta$, SII) in the Gemini/GMOS spectrum (Figure~\ref{fig:HostGalSpectrum}) as well as the detection of this source in 1550 seconds of Galaxy Evolution Explorer \citep[GALEX;][]{Morrissey+2007, Bianchi+2017} near-UV data (21.2 mag in NUV) indicate that the galaxy is star-forming (see Section~\ref{sec:sfr}). Deep FUV observations are not available from GALEX for this field (only 95~s of total integration available) but we visually identify a slight excess in photons at the location of the host. 
Within the DECaLS catalog, the galaxy has an $r$-band magnitude $m_r = 17.8$ and is best fit by a S\'ersic profile. 

\subsection{Host Galaxy SED and Spectral Modeling}

We use the Code Investigating GALaxy Emission (\texttt{CIGALE}) package \citep[v. 2025.0; ][]{Boquien+2019} to fit the host galaxy spectral energy distribution (SED) and derive a stellar mass (see Figure \ref{fig:HostGalSED}). Our input photometry spans the UV to far-infrared, and is collated from Panoramic Survey Telescope and Rapid Response System \citep[Pan-STARRS;][]{chambers2016pan}, Sloan Digital Sky Survey \citep[SDSS; ][]{2000AJ....120.1579Y}, DECam Local Volume Exploration Survey \citep[DELVE; ][]{2021ApJS..256....2D}, DECaLS and the Wide-field Infrared Survey Explorer \citep[WISE;][]{wright2010wide} databases.  To correct the photometry for Galactic extinction, we assume $R_V = 3.1$ and query the IRSA dust extinction database\footnote{https://irsa.ipac.caltech.edu/applications/DUST/} at the FRB position and find $\rm E(B-V) = 0.02$ \citep{SchlaflyFinkbeiner2011}. We incorporate the filter transmission curves for each band from the Spanish Virtual Observatory (SVO) filter database for our extinction correction calculations \citep{Rodrigo+2020, Rodrigo+2024}. We additionally convert all magnitudes to the AB system where applicable. Following the methodology of \citet{Simha+2020}, we assume a delayed-exponential star-formation history (SFH), a synthetic stellar population prescribed by \citet{bc03}, the
\citet{Chabrier03} initial mass function (IMF), dust attenuation models adapted
from \citet{calzetti01}, and infra-red emission templates from \citet{dale14},  where the active galactic nucleus fraction to the 5-20 $\rm \mu m$ emission was capped at 20\%. 
For the host galaxy of FRB 20241228A, we find a best fit stellar mass of ($2.6 \pm 0.3) \times 10^{10} ~M_\odot$ (68\% confidence intervals). 

\subsection{Star-formation rate estimate}
\label{sec:sfr}
To estimate line fluxes from the GMOS spectrum and star formation rate (SFR) of the host, we use the \texttt{pPXF} package \citep{Cappellari2017} to fit the main nebular emission lines. 
We correct the line fluxes for Galactic reddening with the same method as previously explained for the photometry. We tabulate the extinction-corrected line fluxes in Table \ref{tab:line_fluxes}. Based on the observed $\rm H\beta/H\alpha$ flux ratio, and assuming the intrinsic atomic emission flux ratio of 2.87 \citep{OsterbrockFerland2006}, we estimate the intrinsic reddening from the host galaxy of $A_V = 1.77$~mag. Subsequently, we estimate the $\rm H\alpha$ luminosity as $(5.79\pm0.03) \times10^{41}~\rm erg~s^{-1}$. Assuming the \citet{Kennicutt1998} relation for inferring SFR from $\rm H\alpha$ flux, and multiplying by 0.63 to convert to the Chabrier IMF \citep{Conroy+2009}, we estimate the current host galaxy SFR is $2.9\pm0.8 \rm~M_\odot/yr$ (30\% scatter in the Kennicutt relation).

\begin{deluxetable}{cc}
\tabletypesize{\normalsize} 
\tablewidth{0pt}
\tablecaption{Nebular Emission Line Fluxes for the Host of FRB 20241228A
\label{tab:line_fluxes}}
\tablehead{
    \colhead{Emission line} & \colhead{Flux}\\
    \colhead{} & \colhead{($\rm 10^{-16}~erg~s^{-1}~cm^{-2}$)}
}
\startdata
     $\rm H\alpha$ & $20.8 \pm 0.1$\\
     $\rm H\beta$ & $3.94 \pm 0.08$\\
     $\rm [NII]6583$ & $8.3 \pm 0.1$\\
     $\rm [SII]6716$ & $1.61 \pm 0.06$\\
     $\rm [SII]6731$ & $20.8 \pm 0.1$
\enddata
\tablecomments{[OIII] emission features were not detected from this galaxy and the [OII] doublet is outside the wavelength coverage. The reported fluxes are corrected for Milky Way dust extinction.}
\end{deluxetable}

\subsection{Galactocentric Offset}

Following the method presented by \citet{Gordon2025}, we determine the FRB’s offset from the host galaxy center by calculating the distance from each point in the FRB localization probability density function (PDF) to the galaxy center, weighted by the PDF value at that point. We find the FRB has an average galactocentric host galaxy offset of $46^{+47}_{-32}$ kpc, in which the median and uncertainties have been calculated through weighting by the localization probability at every point within the ellipse. To determine the angular size of the host galaxy, we use the \texttt{GALFIT} software package \citep{Peng2002}. We fit an elliptical isophote to the DECaLS $r$-band image of the host. The best-fit model gives a half-light radius of $2.3'' \pm 0.2''$. At a redshift of $z = 0.1614$, this corresponds to a physical size of $6.6\pm0.6$ kpc, assuming cosmological parameters from \citet{planck2018vi}. In combination with the offset, this corresponds to a host-normalized offset (e.g., normalized by the size of the host) of $7^{+7}_{-5}r_e$ (where $r_e$ is the effective radius). We caution that the uncertainties are very large here (although is supportive of a non-zero offset) and hence we do not further interpret this offset.

\section{Burst Properties} 
\label{sec: burst properties}
\subsection{Morphology}

We follow the method outlined by \citet{scm+25} and \citet{csp+24} to determine the morphological burst properties of FRB 20241228A. In particular, we employ the \texttt{fitburst} fitting algorithm which assumes a given burst can be described using a set of exponentially-modified Gaussians \citep{fonseca_fitburst}. For each Gaussian sub-burst, \texttt{fitburst} performs a joint 2D spectro-temporal fit that includes the burst's scattering timescale and DM, along with a per sub-burst spectral index, spectral running, and width. In Table \ref{tab:burst_props}, we list the structure-maximizing DM \citep{smp19}, as well as the \texttt{fitburst}-determined sub-burst widths, scattering timescale, and bandwidth for FRB 20241228A. We also determine the peak flux, specific fluence, and radio luminosity \citep[calculated using the source's redshift and cosmological parameters from ][]{planck2018vi} for the burst by using its baseband position to correct for CHIME's instrumental beam effects  \citep{aaa+23_basecat}.

\begin{deluxetable*}{l c}
\tabletypesize{\normalsize}
\tablewidth{0pt}
\tablecaption{Burst Properties
\label{tab:burst_props}}
\tablehead{
    \colhead{Property} & \colhead{Burst A}
}
\startdata
Time of Arrival (UTC)          & 2024-12-28 15:55:56.984 \\
Real-time Signal-to-Noise          & 50 \\
Baseband Signal-to-Noise          & 202 \\
Dispersion Measure       & $246.53 \pm 0.02$ pc cm$^{-3}$ \\
Width                         & 1: $0.59 \pm 0.01$ ms \\
                              & 2: $0.533 \pm 0.005$ ms \\
Scattering Timescale  & $1.05 \pm 0.01$ ms \\
Peak Flux                     & $41 \pm 4$ Jy \\
Fluence                       & $140 \pm 14$ Jy ms \\
Luminosity                    & $3.7 \pm 0.4 \times 10^{43}$ erg s$^{-1}$\\
Bandwidth                     & 400 MHz \\
Scintillation BW              & 18 kHz \\
\enddata
\tablecomments{Time of arrival (ToA; topocentric) is calculated with the baseband data at 400.0 MHz using a DM constant of $1.0 / (2.41 \times 10^{-4}$ MHz$^2$ s / (pc cm$^-3$). The S/N corresponds to the real-time detection S/N at a time resolution of 0.983 ms. The baseband S/N is calculated at a time resolution of 0.655 ms. The scintillation bandwidth is measured at 600 MHz.}
\end{deluxetable*}

\subsection{Scintillation}
Following the method outlined by \citet{npb2025}, we search for evidence of scintillation within the spectrum of FRB 20241228A. As a brief overview of the scintillation pipeline, we first upchannelize the data in frequency by performing a fast Fourier transform (FFT) to switch between the time and frequency domains. We search over a range of FFT upchannelization factors, ranging from a user-determined minimum (typically set to 16) to a maximum of 1/total burst width. For each upchannelization factor, we sub-band the data and examine the autocorrelation function (ACF), masking the natural zero-lag spike and normalizing such that the peak of the ACF corresponds to the square root of the modulation index. Due to reflections between CHIME's reflectors and feed lines, there is a 30-MHz frequency scale in the ACF that is not astrophysical. Additionally, there can be a leftover upchannelization artifact which occurs at 0.39 MHz. We do not attempt to remove the 30-MHz ripple, meaning we lose sensitivity to scintillation scales close to 30\,MHz. We do fit a model to remove the leftover upchannelization artifact. For each sub-band, we fit a Lorentzian where the half-width at half-maximum corresponds to the scintillation bandwidth. To convince ourselves that the frequency scale we measure in the ACF is from astrophysical scintillation, we search for a frequency-dependent variation in this scale, $\Delta\nu_{\textrm{DC}} \propto \nu^\alpha$, that is consistent with an exponential with $\alpha \approx 4$, as expected for scintillation from a screen comprised of Gaussian phase fluctuations\footnote{In more complex scattering configurations, the scaling of $\nu$ can be more complicated, with significantly shallower slopes in cases where one screen is resolving another.}.

\begin{figure}
    \centering
    \includegraphics[width=\columnwidth]{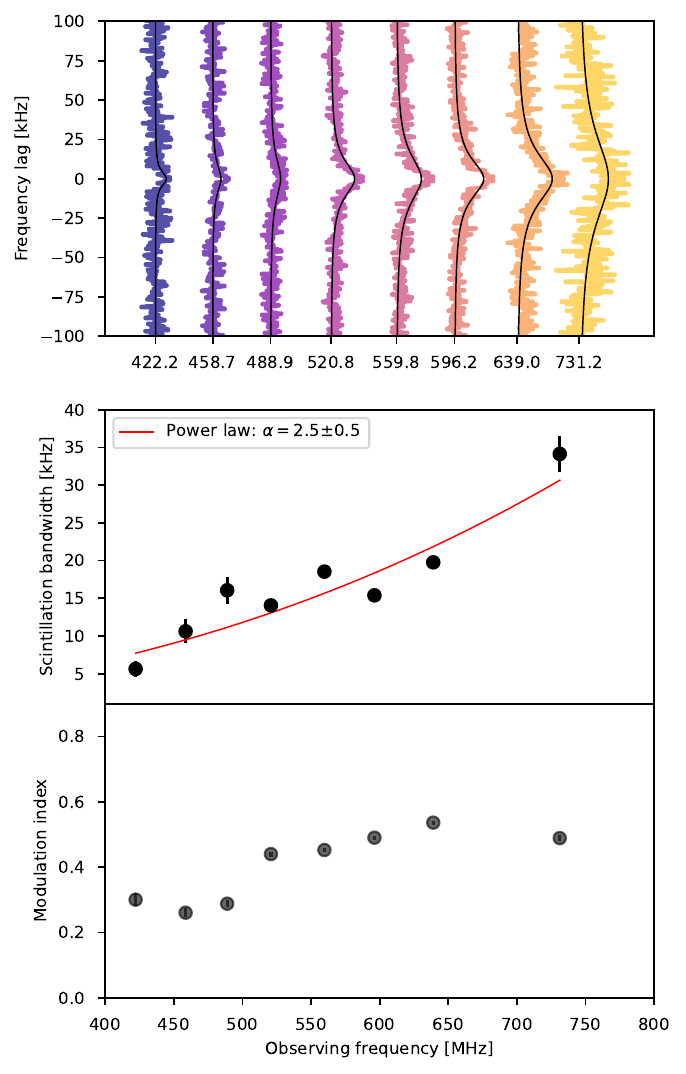} 
    \caption{\textit{Top panel}: Autocorrelation function for eight different subbands in the $400-800$ MHz CHIME/FRB band for FRB 20241228A. The black curve shows the best fit Lorentzian with the scintillation bandwidth corresponding to the FWHM of this function. We do not subtract the 30 MHz ripple from CHIME before calculating the ACF. However, it is much larger the kHz structure probed here. This panel shares the same x-axis as the other two panels, but with different tick marks. \textit{Middle panel}: Scintillation bandwidths from the Lorentzian fits as a function of frequency. The best fit power-law is shown as a red line, and is consistent (within 2$\sigma$ with the expected $\alpha = 4$). \textit{Bottom panel}: Modulation index (proportional to height of ACF at zero lag) across the eight sub-bands. We caution that by not removing the 30 MHz ripple prior to the ACF, the absolute magnitude of the modulation indices may be affected. }
    \label{fig:ACF}
\end{figure}

For FRB~20241228A, we find the scintillation is most clearly visible at an upchannelization factor of 1024 (and hence frequency resolution of 0.7 kHz). We sub-band the 400 to 800-MHz CHIME bandwidth into eight distinct sub-bands, with equal S/N in each band. We then determine a per-subband ACF with a Lorentzian fit (see top panel of Figure \ref{fig:ACF}). The frequency scale observed in the ACF clearly evolves with frequency (Figure\,\ref{fig:ACF}). We measure $\alpha=2.8\pm0.7$, consistent within 2$\sigma$ with the expectation of $\alpha \approx 4$ (see middle panel of Figure \ref{fig:ACF}), thus confirming this frequency scale to be astrophysical scintillation. We measure a characteristic scintillation bandwidth of $18 \pm 3$ kHz at 600 MHz. 

The expected scintillation bandwidth from the Milky Way (MW) in this direction (Galactic latitude $\sim$ 3.5\degree, Galactic longitude $\sim$ 62.5\degree) is $\sim200$ kHz at 600 MHz \citep{ne2001, NE2001p}. While this is approximately an order of magnitude larger than that seen in our measurements, the NE2001 model can have particularly large uncertainties. The expected scattering timescale from the MW in this direction is 0.0006 ms, three full orders of magnitude lower than our measured scattering timescale. If we assume the scattering and scintillation measurements arise from the same source, then we would expect $2 \pi \tau \Delta \nu_{\textrm{DC}} \approx 1$. Using the $\sim$1 ms measured scattering timescale and 18 kHz scintillation scale, we find $2 \pi \tau \Delta \nu_{\textrm{DC}} \approx 120$, significantly larger than 1. This means that the scattering screen responsible for the 1.05 ms measured scattering in FRB 20241228A is distinct from the 18-kHz scintillation screen along this source's line of sight. While both our scintillation and scattering measurements are discrepant with that predicted by NE2001, the scintillation measurement is in closer agreement (1 order of magnitude) while the scattering measurement is in clear disagreement (3 orders of magnitude). While both screens could be Galactic, this is unlikely as the dominant scattering screen would resolve the scintillation screen, and hence only scintillation features from the scattering screen would be detectable. Hence, we conclude that the scintillation screen is likely Galactic, while the scattering screen is likely extragalactic.

The strength of the scintillation screen is quantified through the modulation index, where in our case the modulation index is equal to the square root of the amplitude of the Lorentzian fit to each ACF. We note that the absolute amplitude of the modulation indices measured here may be complicated by the remnant 30-MHz ripple, as well as any remaining upchannelization artifacts. This will be discussed further below, as well as in future work. Assuming the modulation index is still reliable, we find an index $<1$, suggesting that one screen may be partially resolving the other (see bottom panel of Figure \ref{fig:ACF}). Given that there are two distinct screens along the line of sight to this source (one identified using scattering, the other scintillation), we can constrain the distances to the screen under the assumption that one is resolving the other. We use Eq. 7.6 from \citet{2025arXiv250504576T} with $\nu_{\textrm{DC}} = 18$ kHz at 600 MHz, $\nu = 600$ MHz, and $\tau = 1.05$ ms to derive $d_1 d_2 = \frac{16 \textrm{kpc}^2}{m}$ where $d_1$ is the distance between us and the Galactic screen, $d_2$ is the distance between the extragalactic screen and the FRB, and $m$ is the modulation index. Given the measured modulation index of 0.5, this implies $d_1 d_2 = 32 \textrm{kpc}^2$. 

Using the updated NE2001p model, we estimate the Galactic scintillation screen is located at $d_1 = 0.522$ kpc by searching for the maximum amplitude of the electron density fluctuation power spectrum ($C_n^2$) along the line of sight \citep{ne2001, NE2001p}. In conjunction with our observed modulation index of $0.5$, this implies a distance of $\sim$64 kpc between the FRB and extragalactic screen. We note that this is highly dependent on the estimated Galactic scintillation screen distance. The furthest screen found off the Galactic plane is at a scale height of $\sim$2 kpc \citep{2022MNRAS.515.6198S}. Thus, for a Galactic screen at a distance of 2kpc, the extragalactic screen distance would still be 16 kpc from the FRB. It is possible the FRB is located behind the Galaxy itself or that the scattering screen is located within the circumgalactic medium (CGM). However, we leave further interpretation and investigation of this to future work. 

One important caveat here is that the absolute amplitude of the modulation indices measured here may be complicated both by the remnant 30-MHz ripple, as well as the upchannelization artifacts. Thus, if the source is indeed strongly scintillating and the modulation index is just reduced due to systematics, this would give $d_1 d_2 \lesssim 16$kpc$^2$, implying a distance of $<32$ kpc given $d_2=0.5$kpc or  $<8$ kpc given  $d_2=2$kpc. We again defer further investigation and interpretation of this.

\subsection{Polarimetry}

\noindent We determine the polarimetry of the burst following the pipeline developed by \citet{mmm+21b}. The source has a Faraday rotation measure (RM) of $-38.64 \pm 0.01$ rad m$^{-2}$ with a linear polarization fraction of $85.3 \pm 0.4$\%. The inferred Galactic RM contribution along this line of sight of $8\pm5$ rad m$^{-2}$ \citep{2022A&A...657A..43H}. It does not exhibit any swings in its polarization angle as a function of time (see Figure \ref{fig:polarization} in the Appendix).

\subsection{Rate}
\label{subsec: rate}
\noindent Following the method presented by \citet{chimefrbcatalog1}, we calculate the total exposure within the full-width half-maximum of the CHIME/FRB formed beams at the location of this source to be 119 hours. Using the exposure, source position, and fluence, we calculate a spectral fluence threshold following the method outlined by \citet{jcf+19}. The 68th percentile of the fluence threshold distribution is 18.8 Jy ms, while the 95th percentile is 29 Jy ms. We scale the burst rate, $<0.008$ burst/hr, to a fluence threshold of 5 Jy ms assuming a power-law index of $-1.5$ for the cumulative energy distribution \citep{abb+23} and using the 95th percentile fluence threshold. The upper limit on the burst rate is 0.12 burst/hr above a  fluence threshold of 5 Jy ms.

\section{Discussion and Summary} 
\label{sec: Discussion}

In this work, we provide the localization and burst property analysis for the thus-far non-repeating FRB 20241228A. This source is special as {\it Swift}/XRT was on-target at the source position only 112 seconds after the burst (Tohuvavohu et al. \textit{in prep.}), representing the thus far fastest X-ray trigger for a non-repeating FRB to-date. Tohuvavohu et al. \textit{in prep.} present an upper limit on the persistent X-ray flux of $1.52 \times 10^{-13}$ erg cm$^{-2}$ s$^{-1}$ in the $0.3-10$ keV band 112 seconds after the FRB. Using CHIME/FRB along with the GBO and KKO Outriggers, we obtained a VLBI-level localization for FRB 20241228A and robustly associated it with a host galaxy at a redshift of $z = 0.1614$.  At this redshift, and assuming cosmology from \citet{planck2018vi}, this corresponds to an X-ray luminosity limit of $1.2 \times 10^{43}$ erg s$^{-1}$. Given the radio luminosity presented in Table \ref{tab:burst_props}, this corresponds to  $L_x / L_r < 1$. The implications of this are discussed in detail in Tohuvavohu et al. \textit{in prep.}.

In this work, we explore the host galaxy and burst properties of FRB 20241228A. It is found in a spiral star-forming galaxy, as is consistent with that found for most FRB host galaxies \citep[e.g.][]{bmk+23}. Indeed, while the measured SFR is on the higher end when compared to those in \citet{Gordon+2023}, the stellar mass is also commensurately high, leading to an sSFR right at the median value of the full sample. 
Thus, overall the host galaxy properties of FRB 20241228A resemble those of the broader FRB host population.

We did not find any other bursts within the CHIME/FRB database at the location of FRB 20241228A. The source's full 400-MHz bandwidth closely resembles that of the non-repeating FRB population, further supporting its classification as a non-repeater \citep{ple21, scm+24}.
However, the upper limit on the source's rate (0.12 burst/hr) is consistent with that of most repeating FRBs detected by CHIME/FRB \citep{abb+23, csp+24} and the source's duration ($\sim$1 ms) is consistent with both the repeating and non-repeating populations reported in \citet{scm+24} and \citet{csp+24}. Additionally, the lifetime of FRBs is highly uncertain, and thus FRB 20241228A may just be entering its active phase. 

We measure both scattering and scintillation from two distinct plasma screens along the line of sight to FRB 20241228A. We suggest that the scattering screen is located in the host galaxy, while the scintillation screen is in the Milky Way. The modulation index for the scintillation is $<1$, possibly suggesting one screen is resolving the other, and hence placing the extragalactic screen $>10$ kpc from the FRB. However, we leave further investigation and interpretation of this result to future work.

FRB 20241228A represents a unique and exciting example of a new era in FRB-science: coordinated, rapid X-ray follow-up and sub-arcsecond VLBI localizations. With the addition of the third CHIME/FRB Outrigger, we will be able to pinpoint FRBs on $\sim$50-mas-level scales. Combined with deep, rapid X-ray observations, this capability will enable stringent constraints on FRB progenitor models and their local environments.

\begin{acknowledgments}
We acknowledge that CHIME and the \kkoname{} Outrigger (KKO) are built on the traditional, ancestral, and unceded territory of the Syilx Okanagan people. \kkonamecaps{} is situated on land leased from the Imperial Metals Corporation. We are grateful to the staff of the Dominion Radio Astrophysical Observatory, which is operated by the National Research Council of Canada. CHIME operations are funded by a grant from the NSERC Alliance Program and by support from McGill University, University of British Columbia, and University of Toronto. CHIME/FRB Outriggers are funded by a grant from the Gordon \& Betty Moore Foundation. We are grateful to Robert Kirshner for early support and encouragement of the CHIME/FRB Outriggers Project, and to Dusan Pejakovic of the Moore Foundation for continued support. CHIME was funded by a grant from the Canada Foundation for Innovation (CFI) 2012 Leading Edge Fund (Project 31170) and by contributions from the provinces of British Columbia, Québec and Ontario. The CHIME/FRB Project was funded by a grant from the CFI 2015 Innovation Fund (Project 33213) and by contributions from the provinces of British Columbia and Québec, and by the Dunlap Institute for Astronomy and Astrophysics at the University of Toronto. Additional support was provided by the Canadian Institute for Advanced Research (CIFAR), the Trottier Space Institute at McGill University, and the University of British Columbia. The CHIME/FRB raw voltage data recording system is funded in part by a CFI John R. Evans Leaders Fund award to IHS.

A.P.C. is a Vanier Canada Graduate Scholar. 
S.S. is supported by the joint Northwestern University and University of Chicago Brinson Fellowship.
K.N. is an MIT Kavli Fellow.
A.M.C. is a Banting Postdoctoral Researcher.
F.A.D. is supported by the Jansky Fellowship.
Y.D. is supported by the National Science Foundation Graduate Research Fellowship under grant No. DGE-2234667. 
W.F. gratefully acknowledges support by the David and Lucile Packard Foundation, the Research Corporation for Science Advancement through Cottrell Scholar Award \#28284, and the NSF (AST-2206494, CAREER grant AST-2047919).
E.F. is supported by the National Science Foundation (NSF) under grant number AST-2407399.
J.W.T.H. and the AstroFlash research group acknowledge support from a Canada Excellence Research Chair in Transient Astrophysics (CERC-2022-00009); an Advanced Grant from the European Research Council (ERC) under the European Union’s Horizon 2020 research and innovation programme (`EuroFlash'; Grant agreement No. 101098079); and an NWO-Vici grant (`AstroFlash'; VI.C.192.045).
V.M.K. holds the Lorne Trottier Chair in Astrophysics \& Cosmology, a Distinguished James McGill Professorship, and receives support from an NSERC Discovery grant (RGPIN 228738-13).
C. L. acknowledges support from the Miller Institute for Basic Research at UC Berkeley.
K.W.M. holds the Adam J. Burgasser Chair in Astrophysics and is supported by NSF grant 2018490.
D.M. acknowledges support from the French government under the France 2030 investment plan, as part of the Initiative d'Excellence d'Aix-Marseille Universit\'e -- A*MIDEX (AMX-23-CEI-088).
M.N. is a Fonds de Recherche du Quebec -- Nature et Technologies~(FRQNT) postdoctoral fellow.
A.P. is funded by the NSERC Canada Graduate Scholarships -- Doctoral program.
A.B.P. is a Banting Fellow, a McGill Space Institute~(MSI) Fellow, and a Fonds de Recherche du Quebec -- Nature et Technologies~(FRQNT) postdoctoral fellow.
Z.P. is supported by an NWO Veni fellowship (VI.Veni.222.295).
M.W.S. acknowledges support from the Trottier Space Institute Fellowship program.
K.R.S is supported by FRQNT doctoral research award
P.S. acknowledges the support of an NSERC Discovery Grant (RGPIN-2024-06266).
V.S. is supported by FRQNT doctoral research award
K.S. is supported by the NSF Graduate Research Fellowship Program.
\end{acknowledgments}


\facilities{CHIME, KKO, GBO, Gemini
(GMOS)}
\software{\texttt{Fitburst}, \texttt{PATH}, \texttt{ffff-pz}, \texttt{Marz}, \texttt{CIGALE}}

\bibliography{frbrefs}
\bibliographystyle{aasjournalv7}



\appendix 

\begin{figure}
    \centering
    \includegraphics[width=0.3
    \linewidth]{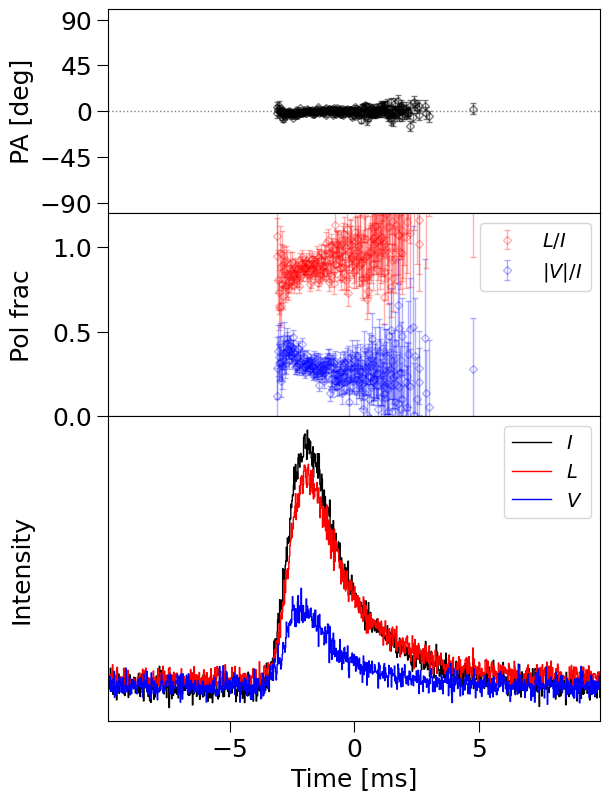}
    \caption{Fractional linear and circular polarization for FRB 20241228A. We do not show the first, weaker sub-burst as the linearly polarized S/N is less than 5. \textit{Top panel}: Polarization angle (PA) over the primary sub-burst of FRB 20241228A. The burst is de-dispersed to the structure maximizing DM presented in Table \ref{tab:burst_props}, and the data is at a time resolution of 81.92 $\mu$s. \textit{Middle panel:} Linear (L/I) and circular ($|$V$|$/I) polarization fractions. \textit{Bottom panel:} Linear and circular polarization profile as compared to the full intensity profile.}
    \label{fig:polarization}
\end{figure}

\noindent In Figure \ref{fig:polarization}, we show the polarization for the primary sub-burst of FRB 20241228A. 

\begin{figure*}
    \centering
    {\includegraphics[width=0.8\textwidth]{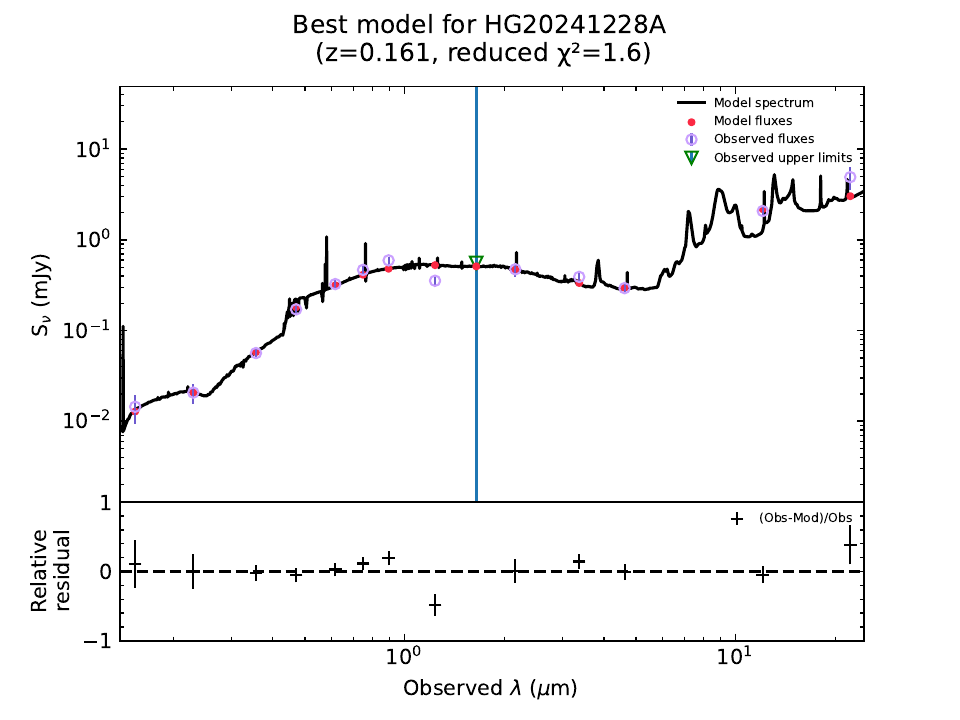} }
    \caption{SED fit for the host galaxy photometry (purple circles with errorbars) using CIGALE. The black line shows the best model fit and is composed of various model components (colored lines). The input photometry is from the Pan-STARRS, SDSS, DELVE, DECaLS, and WISE databases. See Table \ref{tab:photometry} for photometry values.
    }
    \label{fig:HostGalSED}
\end{figure*}

\noindent In Table \ref{tab:photometry}, we list the photometric surveys used to create the SED shown in Figure \ref{fig:HostGalSED} for the host galaxy of FRB 20241228A.

\begin{table}
    \centering
    \begin{tabular}{|c|c|c|}
\hline
Survey & Band &  Mag (AB) \\
\hline
  GALEX &  FUV &  $21.0 \pm 0.4$ \\
   &  NUV &  $20.6 \pm 0.2$ \\
  \hline
  SDSS &    u &   $19.52 \pm 0.06$ \\
   &    g &   $18.32 \pm 0.01$ \\
   &    r &   $17.619 \pm 0.009$ \\
   &    i &   $17.228 \pm 0.009$ \\
   &    z &   $16.96 \pm 0.02$ \\
  \hline
 2MASS &    J &   $17.5 \pm 0.1$ \\
  &    H &   $> 17$  \\
  &    K &   $17.2 \pm 0.2$ \\
 \hline
  WISE &   W1 &   $17.42 \pm 0.03$ \\
   &   W2 &   $17.73 \pm 0.06$ \\
   &   W3 &   $15.60 \pm 0.08$ \\
   &   W4 &   $14.7 \pm 0.3$ \\
\hline
\end{tabular}
\caption{Extinction-corrected optical photometry of the host galaxy obtained from GALEX, SDSS, 2MASS and WISE.}
    \label{tab:photometry}
\end{table}

\end{CJK*}
\end{document}